# Terahertz wave generation using a soliton microcomb

**SHUANGYOU ZHANG,[1] JONATHAN M. SILVER,[1,2] XIAOBANG SHANG,[1] LEONARDO DEL BINO,[1,3] NICK M. RIDLER,[1] AND PASCAL DEL'HAYE[1,*]**

[1]*National Physical Laboratory (NPL), Teddington, TW11 0LW, United Kingdom*
[2]*City, University of London, London, EC1V 0HB, United Kingdom*
[3]*Heriot-Watt University, Edinburgh, EH14 4AS, Scotland*
**pascal.delhaye@npl.co.uk*

**Abstract:** The Terahertz or millimeter wave frequency band (300 GHz - 3 THz) is spectrally located between microwaves and infrared light and has attracted significant interest for applications in broadband wireless communications, space-borne radiometers for Earth remote sensing, astrophysics, and imaging. In particular optically generated THz waves are of high interest for low-noise signal generation. Here, we propose and demonstrate stabilized terahertz wave generation using a microresonator-based frequency comb (microcomb). A unitravelling-carrier photodiode (UTC-PD) converts low-noise optical soliton pulses from the microcomb to a terahertz wave at the soliton's repetition rate (331 GHz). With a free-running microcomb, the Allan deviation of the Terahertz signal is $4.5\times10^{-9}$ at 1 s measurement time with a phase noise of -72 dBc/Hz (-118 dBc/Hz) at 10 kHz (10 MHz) offset frequency. By locking the repetition rate to an in-house hydrogen maser, in-loop fractional frequency stabilities of $9.6\times10^{-15}$ and $1.9\times10^{-17}$ are obtained at averaging times of 1 s and 2000 s respectively, limited by the maser reference signal. Moreover, the terahertz signal is successfully used to perform a proof-of-principle demonstration of terahertz imaging of peanuts. Combining the monolithically integrated UTC-PD with an on-chip microcomb, the demonstrated technique could provide a route towards highly stable continuous terahertz wave generation in chip-scale packages for out-of-the-lab applications. In particular, such systems would be useful as compact tools for high-capacity wireless communication, spectroscopy, imaging, remote sensing, and astrophysical applications.

## 1. Introduction

Low noise microwave and terahertz-wave signals are widely used in many areas, such as radar, wireless communication, and global navigation satellite systems. Photonic signal generation methods have been demonstrated with bandwidths up to a few THz [1-5], as well as with ultra-low phase-noise, using optoelectronic oscillators [6], electro-optical-frequency division [7], Brillouin oscillators [8], and optical frequency division based on optical frequency combs [9-11]. Among these different photonic methods, in particular optical frequency division based on optical frequency combs enables microwave signal generation with sub-fs timing jitter [9, 10]. Ultralow noise THz radiation is usually generated by photomixing two optical waves which are phase locked to an optical frequency comb [2, 12-14]. However, the excellent performance comes at the expense of size and complexity, which makes it difficult to use in out-of-the-lab applications.

As an alternative platform, microresonator-based frequency combs ("microcombs") based on parametric four-wave mixing in monolithic high-Q microresonators provide a promising scheme to miniaturize optical frequency comb systems [15, 16]. In particular, soliton formation in microresonators has been demonstrated as a source for low noise, fully coherent frequency combs [17-19]. Recently, low noise soliton microcombs have been generated in a fully integrated and battery-powered system [20] and with sub-mW optical pump power [21].

Compared with conventional mode-locked frequency combs, microcombs have the potential for chip-level integration and low power consumption. Moreover, because of the small footprint, repetition rates ($f_{rep}$) of microcombs from several GHz up to 1 THz can be easily realized. Several demonstrations of microwave signal generation have been made with microcombs [22-29]. High spectral purity microwave generation based on soliton microcombs has been demonstrated with phase noise much lower than existing radio frequency photonic oscillators of similar size, weight, and power consumption [25]. Recently, on-chip high-power THz wave generation was demonstrated using parametric sideband generation in a microresonator with frequency stability of $6\times10^{-10}$ at 100 s [30].

In this Letter, by using a monolithically integrated UTC-PD, we propose and demonstrate an ultrastable continuous terahertz wave generation scheme with a soliton microcomb. The soliton microcomb with a single soliton circulating inside the cavity is generated in a 200-μm-diameter silica microtoroid with an $f_{rep}$ of 331 GHz. The UTC-PD is used to directly convert the optical pulse train to a continuous THz wave with ~ -10 dBm power. The generated THz signal has a frequency stability of $4.5\times10^{-9}$ at 1 s integration time and a phase noise of -72 dBc/Hz (-118 dBc/Hz) at 10 kHz (10 MHz) offset frequency when the microcomb is free-running. It can be significantly improved to have the same frequency stability ($2\times10^{-13}$ at 1 s integration time) as a hydrogen maser when locking $f_{rep}$ to the maser. The demonstrated technique could meet the demand for portable and compact THz wave generators with ultralow noise and stable operation at room temperature [31], and has the potential to be widely used for THz metrology, imaging, and wireless communication [2, 4, 32].

## 2. Experimental setup

Figure 1(a) shows the schematic of the experimental setup for the THz wave generation based on a soliton microcomb. A 1.5 μm wavelength external cavity diode laser (ECDL) with a short-term linewidth of <10 kHz is used as the pump laser for generating a soliton microcomb while another ECDL at 1.3 μm is used as an auxiliary laser to passively stabilize the circulating optical power inside the microresonator, and hence the temperature of the microresonator [21]. Figure 1(b) shows the 200-μm-diameter fused silica microtoroid used in the experiments with a free spectral range of 331 GHz [33]. It is fabricated from a silicon wafer with a 6-μm layer of silicon dioxide ($SiO_2$). A mode family of the resonator with a quality factor of ~ $4\times10^{8}$ is chosen for soliton generation. The 1.5-μm pump laser is tuned to a wavelength around 1554 nm to generate the corresponding soliton microcomb. Another optical mode at 1336 nm with a quality factor of ~ $3\times10^{8}$ is used as an auxiliary mode to compensate the temperature variation of the microresonator during soliton generation. These two lasers are combined with a wavelength division multiplexer (WDM) and evanescently coupled into the microresonator via a tapered optical fiber. Two fiber polarization controllers (PCs) are used to match the polarizations of the two lasers. At the resonator output, another WDM is used to separate the auxiliary laser from the generated comb light. One part of the comb light is sent to an optical spectrum analyzer (OSA) for monitoring, and the rest is filtered by a tunable fiber Bragg grating (FBG) notch filter to suppress the pump light. One part of the filtered comb light is monitored by a low-speed photodiode (PD1 in Fig. 1(a)) for monitoring the comb power, and the rest is amplified by an Erbium-doped fiber amplifier (EDFA) and detected with a J-band UTC-PD for converting the $f_{rep}$ signal of the generated soliton microcomb into a THz wave. The optical power of the auxiliary light is monitored by a third photodiode (not shown in Fig. 1(a)). The frequency of the auxiliary laser is tuned into its resonance from the blue detuned side and fixed on the blue side of the resonance. Note that, the auxiliary laser is free-running without feedback control on the laser frequency or the power during the soliton generation. By optimizing the laser detuning and optical power of the auxiliary laser, soliton states can be accessed by slowly tuning the pump laser frequency into a soliton regime from the blue detuned side. A detailed description of the soliton generation process can be found in the reference [21].

In the experiments, 100 mW pump light is used to generate the soliton frequency comb while ~50 mW of auxiliary laser light is used to passively compensate thermally induced resonance shifts of the resonator. Figure 1(c) shows a typical optical spectrum of the single-soliton microcomb used to generate the THz wave. The 3-dB bandwidth of the spectrum is around 4.7 THz, corresponding to a sub-100 fs optical pulse.

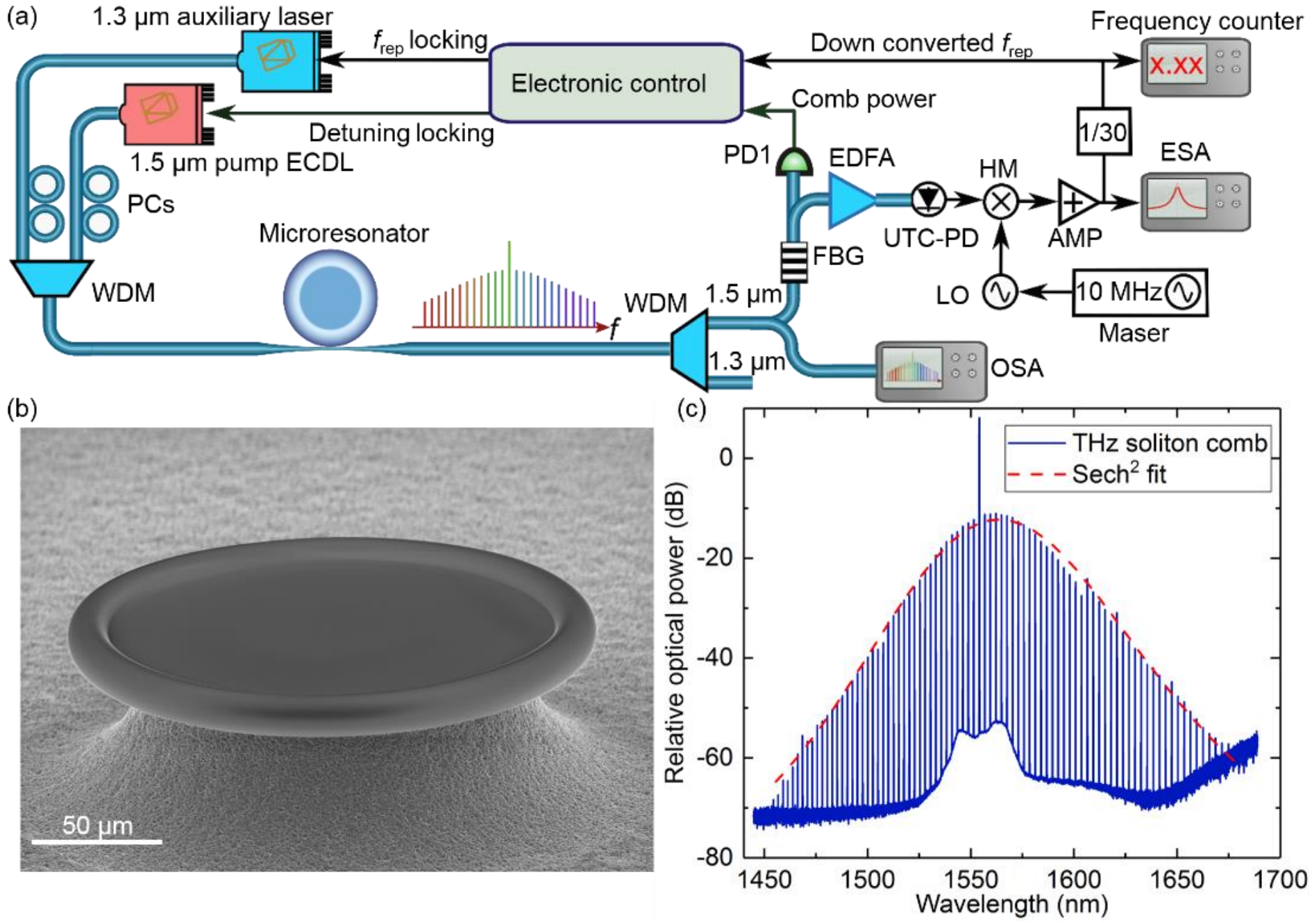

Fig. 1. Experimental setup and THz soliton frequency comb spectrum. (a) Schematic of the setup for microcomb-based THz wave generation. The 1.3 μm auxiliary laser is used to stabilize the THz signal. ECDL: external cavity diode laser; WDM: wavelength division multiplexer; PC: polarization controller; FBG: fiber Bragg grating; EDFA: Erbium-doped fiber amplifier; UTC-PD: unitravelling-carrier photodiode; PD: photodetector; HM: harmonic mixer; AMP: RF amplifier; LO: local oscillator; OSA: optical spectrum analyzer; ESA: electronic spectrum analyzer. (b) Scanning electron microscope image of a 200-μm-diameter microtoroid. (c) Optical spectrum of the generated single-soliton state pumped with 100 mW optical power and thermally stabilized with ~50mW auxiliary laser power. The generated single soliton frequency comb has a 3-dB optical bandwidth of ~ 4.7 THz, corresponding to a 67 fs optical pulse. The red dashed line shows a fitted sech$^2$ envelope.

## 3. THz wave generation

### *3.1 Free-running THz Wave*

As shown in Fig. 1(a), after sending the frequency comb through the FBG notch filter to suppress the pump light, one part of the filtered comb light is amplified to 40 mW by an EDFA and sent into a UTC-PD. The UTC-PD with a J-band rectangular waveguide port converts the incident optical frequency comb into 331 GHz radiation. After photomixing, the THz signal is guided with a WR-3 waveguide to a harmonic mixer, which is used to down-convert the THz signal to an intermediate frequency (IF) at ~1 GHz for characterizing the THz wave performance. The harmonic mixer operates at the 26th harmonic of the local oscillator (LO) input frequency. The frequency of the LO signal is 12.69 GHz generated from a microwave signal generator (E8257D) referenced to a hydrogen maser. With an average photocurrent of 7 mA, the resulting output power of the generated 331 GHz wave is around -10 dBm. Due to the 40 dB conversion loss from the harmonic mixer, the down-converted IF signal has only -50 dBm RF power and is amplified using low-noise RF amplifiers. Figure 2(a) shows the electronic spectrum of the signal (after down conversion to IF) with a resolution of 1 kHz when the soliton microcomb is free-running. The electronic spectrum has a 2 kHz 3-dB linewidth with >40 dB signal-to-noise ratio (SNR). The red curve in Fig. 2(b) shows the measured power spectral density of the phase noise of the generated THz signal. The single sideband (SSB) phase noise is -72 dBc/Hz at 10 kHz and -85 dBc/Hz at 100 kHz with a noise floor of -118 dBc/Hz (dash line shown in Fig. 2(b)), which is limited by the white noise of RF amplifiers, due to the small input RF power (-50 dBm). We can see that the phase noise measurement between 300 kHz and 2 MHz is limited by the LO microwave reference (purple triangles).

To characterize the frequency stability of the THz signal, the down-converted IF signal is electronically divided by 30, and recorded with a frequency counter. Figure 2(c) shows the Allan deviation of the free-running THz signal with a 1 ms gate time. It is about $4.5\times10^{-9}$ at 1 s measurement time, corresponding to 1.5 kHz fluctuations of the 331 GHz carrier. Figure 2(d) shows the recorded frequency drift of the THz signal. Figure 2(e) depicts the effect of the pump laser frequency on the $f_{rep}$ of the soliton microcomb. The slope is around 46 kHz/MHz. Note that the soliton state remains stable over a 170 MHz pump laser detuning range.

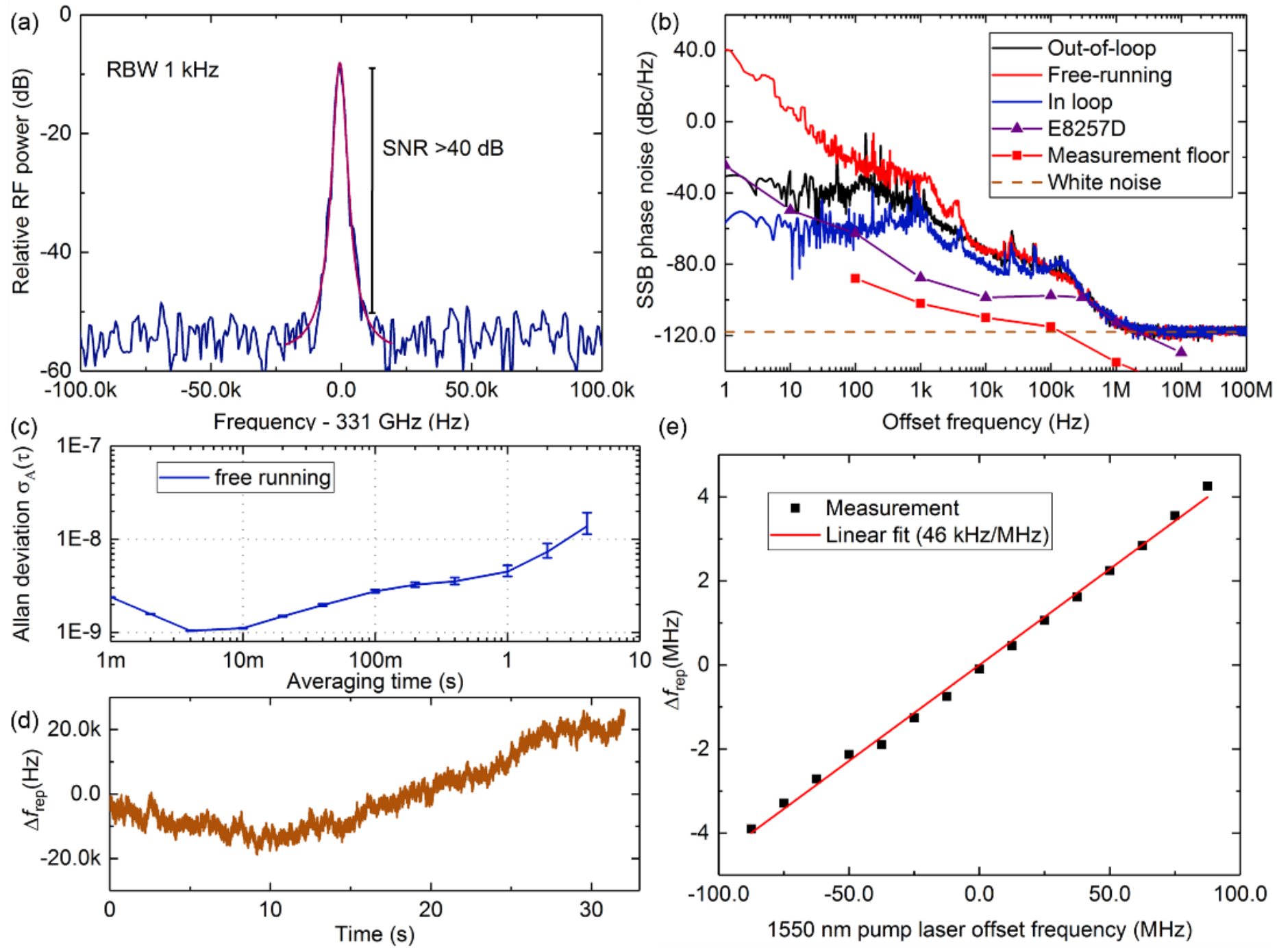

Fig. 2. Performance of the generated THz wave with a free-running soliton microcomb. (a) RF spectrum of the generated THz signal with a 1 kHz resolution bandwidth (RBW). The red line shows a Lorentzian fit. (b) Single sideband (SSB) phase noise spectra of the generated THz signal. The graph shows data for the free-running single-soliton microcomb (red line) and data with frep being stabilized to a hydrogen maser (blue line). The red line with squares is the measurement noise floor of the ESA. The brown dashed line is the white noise floor from the RF amplifier. The purple line with triangles is the phase noise of the LO reference (scaled to 331 GHz). (c) Allan deviation of the free-drifting THz signal. (d) Recorded frequency drift of the THz signal at 1 ms gate time. (e) Frequency dependence of the generated THz signal on the 1550 nm pump laser frequency. The slope is around 46 kHz/MHz.

### *3.2 Stabilizing the THz wave to a hydrogen maser*

To generate a stabilized THz signal, the $f_{rep}$ of the soliton microcomb is phase-locked to the 10 MHz signal from a hydrogen maser via a microwave signal generator and a harmonic mixer (see Fig. 1(a)). Two feedback loops are implemented to ensure the long-term locking of $f_{rep}$, one of the feedback loops stabilizes the pump laser detuning from its resonance via control of the pump laser frequency to ensure long-term operation of the soliton state, and the other one locks $f_{rep}$ via the temperature of the resonator by actuating on the frequency of the auxiliary laser. Due to the dependence of soliton power on pump laser detuning [34], the comb power (detected by PD1 after the FBG) is used to generate an error signal to control the pump laser frequency via the piezo element inside the pump laser. To stabilize $f_{rep}$, the down-converted IF signal (~ 1.04 GHz) is first amplified and electronically divided by 30. Subsequently the divided signal at ~ 34.7 MHz is phase-locked to a maser-stabilized reference signal. The error signal is generated by an analog RF mixer and fed to a proportional-integral (PI) controller. Control of $f_{rep}$ is achieved by actuating on the frequency of the auxiliary laser. The piezo element and pump current of the external cavity auxiliary laser are used for slow and fast control of its frequency, respectively. Modulating the auxiliary laser frequency will change its detuning from its resonance, thereby changing the intracavity auxiliary optical power, and hence the temperature of the resonator. This variation of the temperature changes the overall size and refractive index of the resonator, and hence the value of $f_{rep}$. The effect of the auxiliary optical power on $f_{rep}$ is shown in Fig. 3(a). During these measurements, the optical pump power is kept constant (100 mW) and a microcomb is operated in a single-soliton state while the auxiliary laser power is changed from 40 mW to 130 mW. The pump detuning is locked to a constant setpoint to exclude an additional influence on $f_{rep}$ (46 kHz/MHz as shown in Fig. 2(e)). For stabilization of $f_{rep}$, the operating point of the auxiliary laser power is chosen at around 50 mW. The slope ($\gamma_p$) at this power is about - 500 kHz/mW.

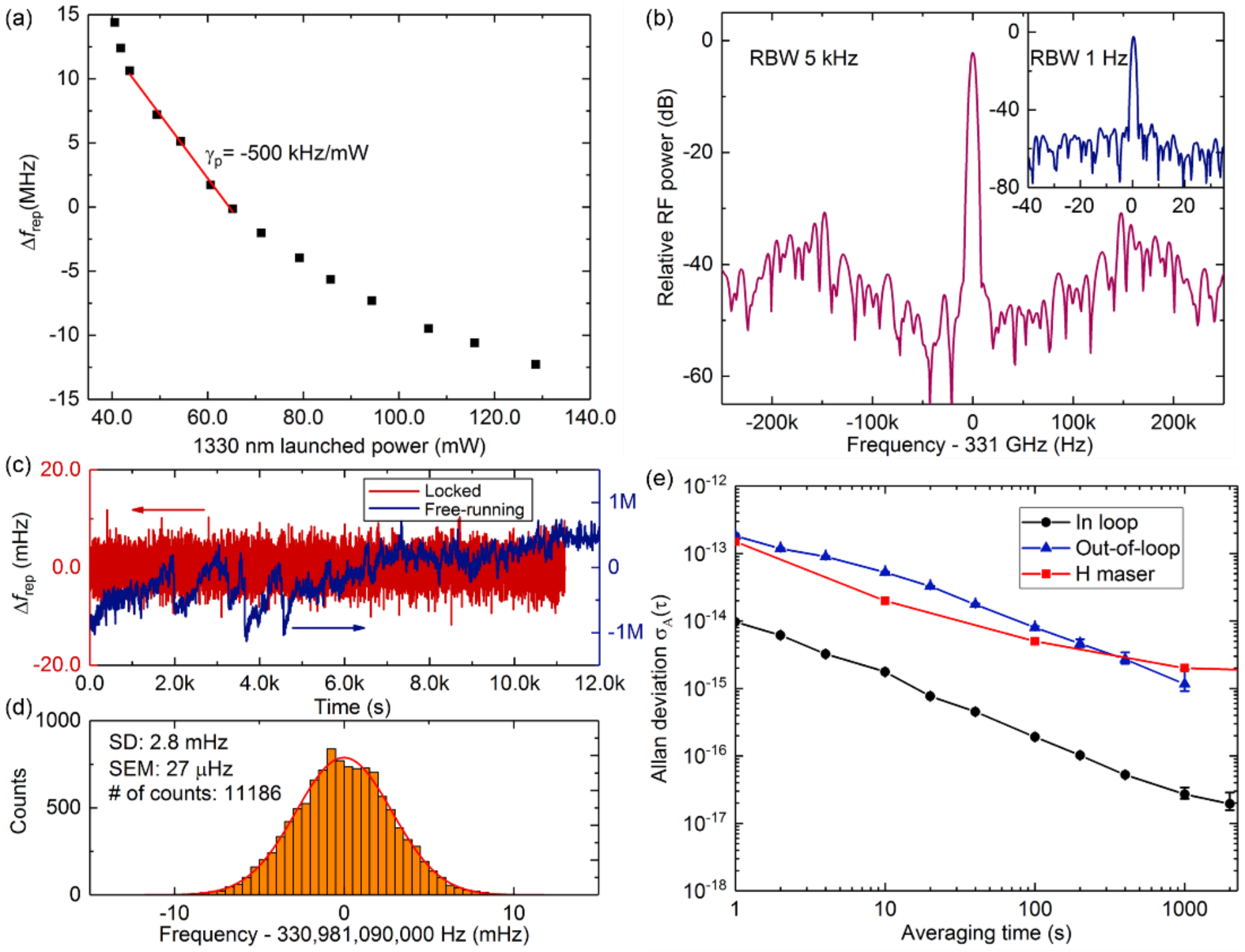

Fig. 3. THz wave stabilized to a hydrogen maser frequency reference. (a) Dependence of the THz signal frequency on the 1330 nm auxiliary power launched into the microresonator. (b) Electronic spectrum of the stabilized THz signal with a 5 kHz RBW. The inset shows the resolution bandwidth limited electronic spectrum (1 Hz RBW). (c) Time series measurement of the variation of the stabilized THz signal (red, left axis) and the free-running THz signal (blue, right axis) at 1 s gate time. (d) Histogram of the stabilized THz signal with a standard deviation (SD) of 2.8 mHz at 1 s gate time and a standard error of the mean (SEM) of 27 μHz. (e) Allan deviations of the in-loop THz signal (black circles), of the out-of-loop THz signal (blue triangles) and of the used hydrogen maser (red squares).

Figure 3(b) shows the electronic spectrum of the THz signal with stabilized $f_{rep}$. We can see that the locking bandwidth is about 150 kHz, which is attributed to the thermal and Kerr response of the resonator. The inset shows a resolution-bandwidth-limited electronic spectrum at 1 Hz resolution bandwidth. The 3-dB linewidth is <1 Hz. The blue curve in Fig. 2(b) shows the in-loop SSB phase noise of the THz signal. The phase noise within the locking bandwidth is well suppressed (~90 dB at 1 Hz offset frequency). A > 3-hour measurement is conducted to evaluate the long-term locking performance. The red curve (left axis) in Fig. 3(c) shows the variation of the locked THz signal while the blue curve (right axis) shows the free drifting $f_{rep}$ for comparison at 1 s gate time. The THz signal is well controlled within 20 mHz for more than 3 hours, whereas without feedback control it drifts over a range of ~1.5 MHz. Figure 3(d) shows the histogram of the stabilized THz signal based on the red curve shown in Fig. 3(c), and it shows a standard deviation of 2.8 mHz. Figure 3(e) shows the Allan deviation for the stabilized in-loop THz signal (black line with circles) while the red line with squares is from the hydrogen maser used in the experiments. The in-loop frequency stability is $9.6\times10^{-15}$ and $1.9\times10^{-17}$ at 1 s and 2000 s averaging time respectively. Note that the Allan deviation of the in-loop THz signal is at least one order of magnitude better than that of the maser used, thus limiting the THz signal stability to the stability of the hydrogen maser.

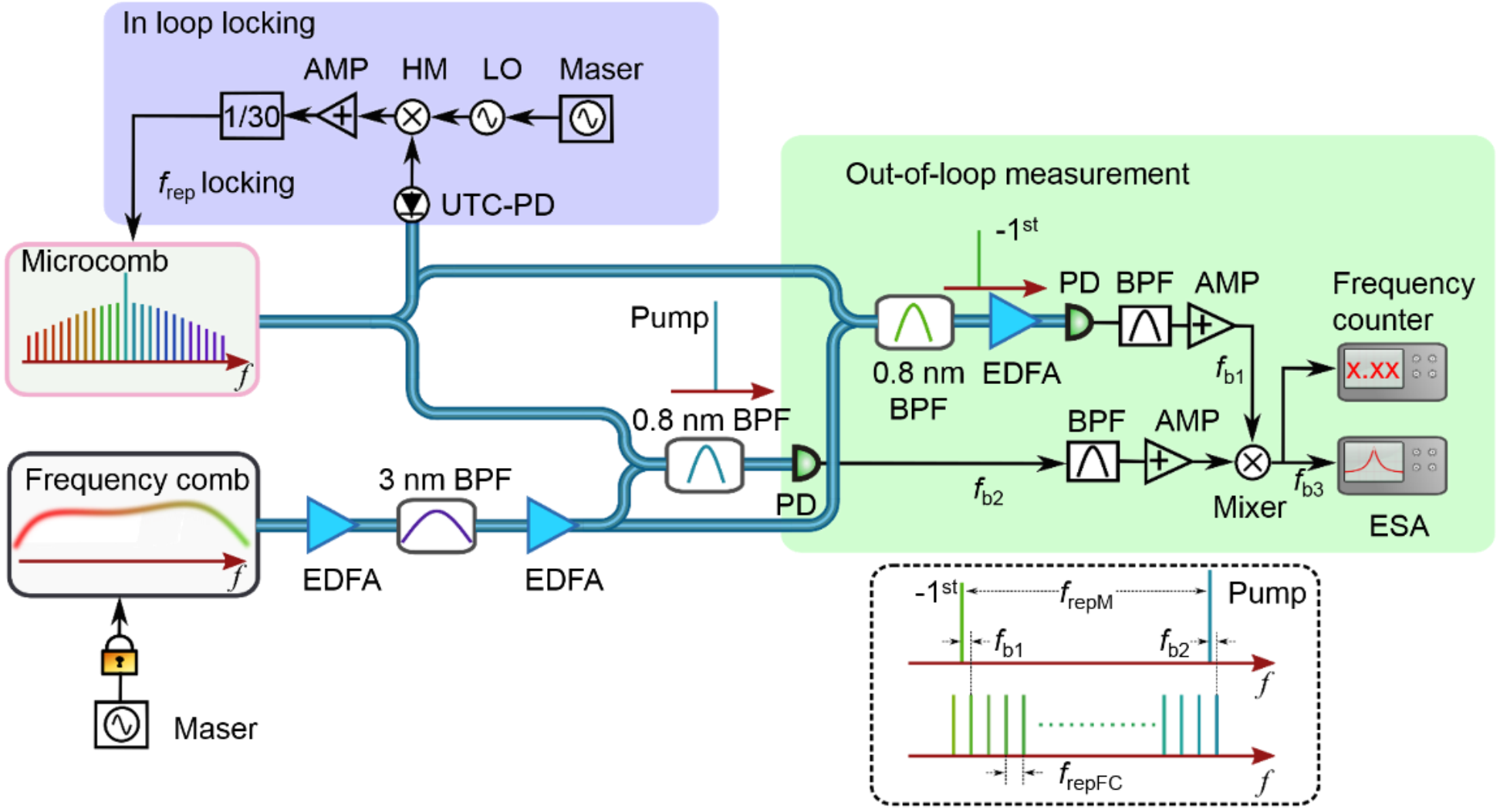

Fig. 4. Experimental setup for out-of-loop measurement of the generated THz wave. $f_{b1}$, $f_{b2}$ are beat note signals between the microcomb (pump mode and first sideband) with a fiber laser reference frequency comb. $f_{repM}$, $f_{repFC}$ are the repetition rates of microcomb and fiber frequency comb, respectively.

### *3.3 Out-of-loop measurement of the THz wave*

To verify the noise performance of the generated THz wave and due to the lack of another UTC-PD in our lab, we compare the frequency stability of $f_{rep}$ of the soliton microcomb against a self-referenced Erbium fiber frequency comb. Figure 4 shows the experimental setup. The out-of-loop measurement is conducted by comparing the pump and the first neighboring microcomb line with their adjacent fiber frequency comb lines, as shown in the dashed box in Fig. 4. To improve the SNR of the optical beat signals between microcomb lines and fiber comb lines, the output of the fiber comb is filtered by a 3-nm optical bandpass filter and amplified by two EDFAs. The microcomb pump mode and first microcomb sideband are selected by two 0.8-nm-bandwidth optical bandpass filter. The selected first microcomb line together with its adjacent fiber comb lines are amplified with another EDFA. After photodetection, the beat signals ($f_{b1}$, $f_{b2}$ as shown in the dashed box in Fig. 4) are electronically filtered and amplified. A mixer is used to generate the frequency difference signal ($f_{b3}$) between $f_{b1}$ and $f_{b2}$. The difference frequency $f_{b3}$ is linked to the difference of the repetition rates of the two combs via $f_{b3} = f_{b1} - f_{b2} = f_{repM} - N f_{repFC}$, where $f_{repM}$, $f_{repFC}$ are repetition rates of soliton microcomb and fiber frequency comb, respectively. N is an integer number and equals to 1323 while $f_{repFC}$ equals to 250 087 558. 9 Hz. In the measurements, $f_{repFC}$ is stabilized to the same hydrogen maser that is used for stabilizing the THz wave and thus $f_{b3}$ can be used to evaluate the noise performance of the generated THz wave. The black curve in Fig. 2(b) shows the out-of-loop SSB phase noise measurement. Compared with the free-running result (red curve in Fig. 2 (b)), the out-of-loop result shows the noise suppression bandwidth only up to 5 kHz, which is limited by the locking bandwidth of the fiber frequency comb. The blue line with triangles in Fig. 3(e) shows the Allan deviation of the out-of-loop THz signal, which is limited by the stability of the hydrogen maser. The small deviation between 1 s and 100 s is attributed to noise from the fiber link and EDFA in the non-common path of the out-of-loop setup.

## 4. Non-destructive THz imaging using soliton-microcomb-based THz source

In a separate proof-of-concept (that does not require stabilization of the soliton comb), we demonstrate non-destructive THz imaging with a microcomb. In future this could enable compact THz sources with a single laser diode integrated with a high-Q microresonator and a UTC-PD instead of using multiple lasers. Figure 5(a) shows the experimental configuration of the THz imaging system using the soliton-microcomb-based THz source. Instead of connecting the UTC-PD to a harmonic mixer, the rectangular waveguide output port is connected to a pyramidal horn antenna for radiating the THz signal. The horn antenna has a gain of 25 dBi with an 11.1 mm × 9.3 mm rectangular aperture. Figure 5(b) shows a photograph of the imaging setup. The THz camera has 1024 pixels (32×32 array) and 1.5 mm spatial resolution. It is placed ~135 mm away from the antenna and in the radiating field of the THz signal. Figures 5(c) and 5(d) show photographs of two peanuts that are used as test samples. The red foam in the photographs is used as a sample holder and is transparent for the 331 GHz signal. The insets in Fig. 5(c) and 5(d) show the internal structures of two peanuts, one with two nuts and the other one with only one nut inside, respectively. Figures 5(e) and 5(f) show their respective THz transmission images. Due to the absorption of the THz wave by the nuts, there are two low-intensity (blue) regions in Fig. 5(e) (two nuts) and only one in Fig. 5(f) (one nut).

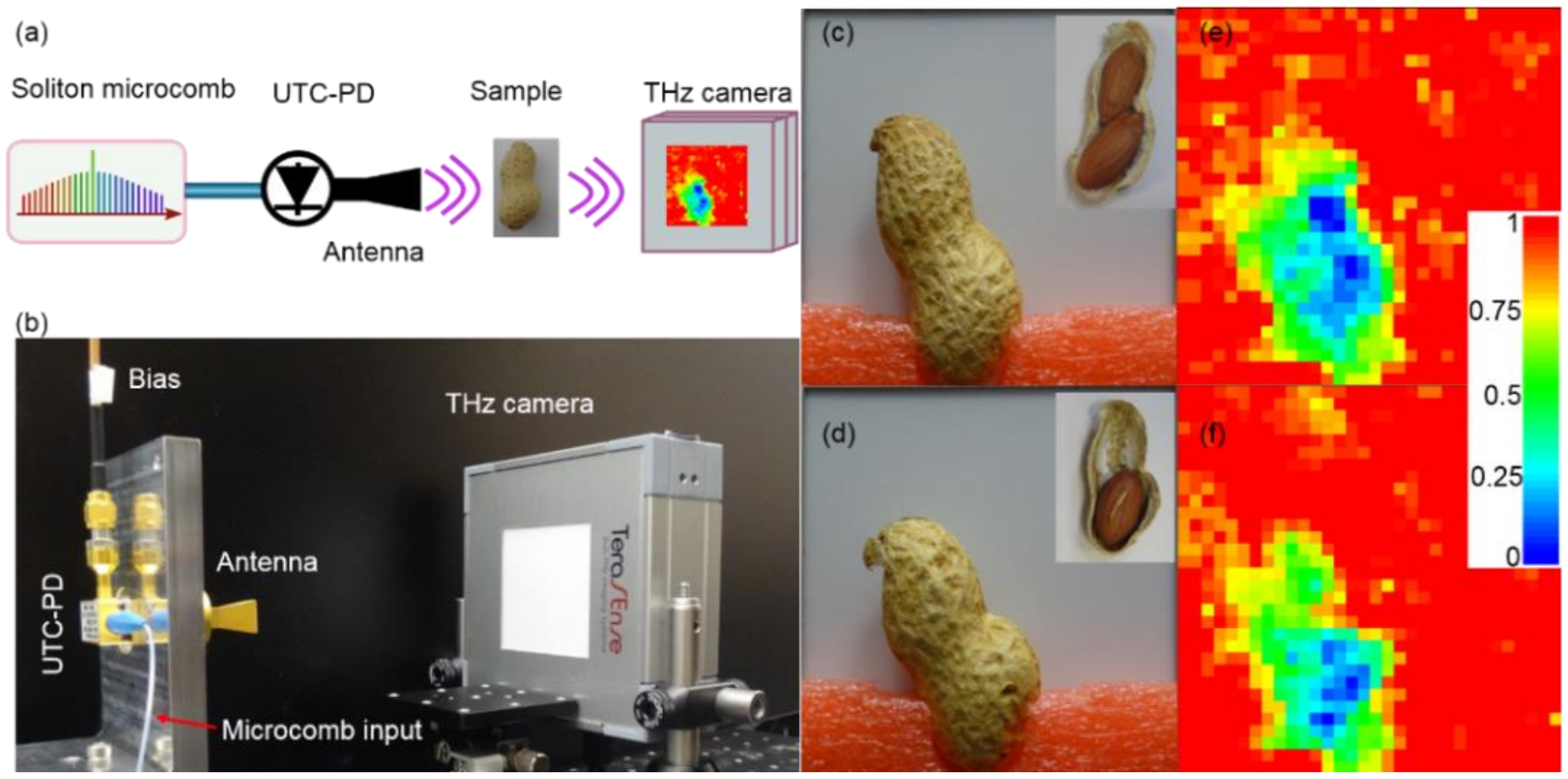

Fig. 5. Non-destructive THz imaging system based on a soliton microcomb. (a) Experimental setup for the THz imaging system using a THz camera. (b) Photograph of the imaging setup. Samples are placed in front of the camera. (c) (d) Photographs of the peanuts with two and one nuts inside, respectively. The insets show the same peanuts without nutshells. The red foam is used as sample holder, and is transparent for the THz wave. (e) (f) THz images of the corresponding peanuts with two nuts (e) and only one nut (f) inside. These THz images are recorded from the THz transmission through the samples. The colour bar shows the THz attenuation ratio.

## 5. Conclusion

In conclusion, we have demonstrated stabilized THz wave signal generation based on a soliton microcomb. The frequency of the THz signal, which is linked to the comb repetition rate, can be controlled with a bandwidth of >150 kHz via the frequency of an auxiliary laser. Using this mechanism for feedback, the THz signal can be phase-locked to a hydrogen maser such that the generated THz wave has the same frequency stability as the maser. Compared with the THz generation schemes based on conventional frequency combs [2, 12], the demonstrated scheme achieves > 40 dB lower phase noise for > 1 MHz offset frequency. It shows a phase noise floor of -118 dBc/Hz. Compared with other signal generation experiments based on soliton mcirocombs [25, 26, 28], our scheme has a 20 dB lower phase noise than these of silica wedge[26] and $Si_3N_4$ resonators [28] and a similar performance to $MgF_2$ resonators [25] when scaling these signals to 331 GHz for comparison. We believe that our stable, atomic-clock-referenced THz signal can be used for THz metrology [32] and THz astrophysics [35, 36], e.g. to search for young galaxies, and to measure the evolution of interstellar matter. Moreover, the demonstrated low-noise THz generation scheme could have a high impact on high-speed THz wireless communication [2], especially for space applications [37, 38], that do not suffer from atmospheric signal attenuation. In addition, we successfully tested the generated THz signal for use in THz imaging applications. THz images were obtained for different samples, non-destructively revealing the internal structure of test samples. By simply changing the diameter of the microresonator, different frequencies of THz waves from 100 GHz to several THz can be generated, which have a wide range of possible applications in agricultural and industrial imaging [39]. By using a high-performance UTC-PD [40], this system could generate milliwatt-level THz signals. Taken together with progress in chip-integrated soliton microcombs [20] and monolithically integrated UTC-PDs [31], these results pave the way for low-noise, room-temperature-operation, chip-based THz wave signal generation, which can be used for high-capacity wireless communication, THz metrology, and non-destructive imaging and sensing.

### *Funding*

Horizon 2020 Marie Sklodowska-Curie Actions (748519, CoLiDR); Horizon 2020 Marie Sklodowska-Curie grant (GA-2015-713694); Horizon 2020 European Research Council (ERC) grant (756966, CounterLight); National Physical Laboratory Strategic Research Programme; Engineering and Physical Sciences Research Council (EPSRC); EPSRC via the CDT for Applied Photonics; JS acknowledges funding from a postdoctoral fellowship of the Royal Academy of Engineering.

### *Acknowledgments*

The authors would like to thank Prof. Stepan Lucyszyn at Imperial College London for the loan of the THz camera used in these experiments. We thank Giuseppe Marra, Ian Hill and Alissa Silva for helpful discussions.